\documentclass[%
nofootinbib,
 aip,
 amsmath,amssymb,
 reprint,%
]{revtex4-1}
\usepackage{graphicx}% Include figure files
\usepackage{dcolumn}% Align table columns on decimal point
\usepackage{bm}% bold math
\usepackage{siunitx}
\usepackage{xcolor}
\usepackage{physics}

\usepackage[utf8]{inputenc}
\usepackage[T1]{fontenc}
\usepackage{mathptmx}
\DeclareUnicodeCharacter{2212}{-}

\begin{document}

\preprint{AIP/123-QED}

\title{High operating temperature plasmonic infrared detectors}
% Force line breaks with \\

\author{L. Nordin}
\thanks{Author to whom correspondence should be addressed: nordin@utexas.edu}
\author{A.J. Muhowski}
\author{D. Wasserman}
\affiliation{ 
Dept. of Electrical and Computer Engineering, University of Texas at Austin, Austin, TX 78758 USA}%

\date{\today}% It is always \today, today,

\begin{abstract}
III-V semiconductor type-II superlattices (T2SLs) are a promising material system with the potential to significantly reduce the dark current of, and thus realize high-performance in, infrared photodetectors at elevated temperatures. However, T2SLs have struggled to meet the performance metrics set by the longstanding infrared detector material of choice, HgCdTe. Recently, epitaxial plasmonic detector architectures have demonstrated T2SL detector performance comparable to HgCdTe in the 77 K - 195 K temperature range. Here we demonstrate a high operating temperature plasmonic T2SL detector architecture with high-performance operation at temperatures accessible with two-stage thermoelectric coolers. Specifically, we demonstrate long-wave infrared plasmonic detectors operating at temperatures as high as 230 K while maintaining dark currents below the ``Rule 07'' heuristic. At a detector operating temperature of 230 K, we realize 22.8\% external quantum efficiency in a detector absorber only 372 nm thick ($\sim\lambda_0 /25$) with peak specific detectivity of \SI{2.29E9}{\cm\Hz^{1/2}\watt^{-1}} at \SI{9.6}{\um}, well above commercial detectors at the same operating temperature.
\end{abstract}

\maketitle

The performance of infrared photodetectors, particularly at long wavelengths, degrades drastically with increasing temperature. Often, expensive and bulky cryo-cooling modules are required to reduce detector dark current and maintain acceptable detector performance. In the long-wave infrared (LWIR $8-\SI{13}{\um}$), detectors must typically be cooled to near liquid nitrogen temperatures in order to achieve reasonable performance metrics, increasing both the cost and weight of infrared sensor systems. The current and long-time state-of-the-art LWIR detector is the HgCdTe (MCT) photodetector.  MCT detectors, in addition to suffering the same high-temperature challenges as any LWIR detector, rely on the epitaxial growth of II-VI alloys, for which uniform growth is notoriously difficult, especially for the high Hg-content HgCdTe alloys required for LWIR detection\cite{Rogalski2019Type-IIPhotodiodes}. Moreover, MCT materials are facing increased restrictions, including being banned in European markets, due to environmental concerns associated with mercury and cadmium \cite{EUDirective}. For this reason there has been historical, and increasing, interest in developing alternative materials and architectures, specifically in the III-V semiconductor family, for LWIR detection\cite{schneider2007quantum,Rogalski2019Type-IIPhotodiodes}.

Type-II superlattices (T2SLs) comprising III-V semiconductor materials offer a promising alternative to the long-dominant MCT material system for infrared detector applications \cite{Smith1987ProposalDetectors,Grein1992MinoritySuperlattices}. Specifically, when compared to the MCT material system, T2SLs are less environmentally problematic, offer more uniform growth (due to the lower vapor pressure of III-V materials), leverage mature III-V semiconductor device fabrication processes, and have been theorized to have significantly lower dark currents\cite{Grein1992MinoritySuperlattices}.  However, T2SL detectors have struggled to meet predicted performance metrics, in part a result of short minority carrier lifetimes \cite{Connelly2010DirectPhotoluminescence} and/or low absorption coefficients \cite{Vurgaftman2016InterbandMaterials}. One increasingly attractive approach to overcoming the poorer performance of T2SL-based detectors has been to leverage detector architectures which enhance the optical fields in the detector \cite{Nolde2015,Goldflam2016, Nolde2019TemperatureAbsorbers, Canedy2019,Letka2019, Nordin_ACSPhot, Abhilasha_GMR, Nordin2021Ultra-thinDetectors, Letka2020} to increase detector absorption while maintaining small detector volumes, and therefore low dark currents. 

\begin{figure}[!t]
  \includegraphics[scale = 0.28]{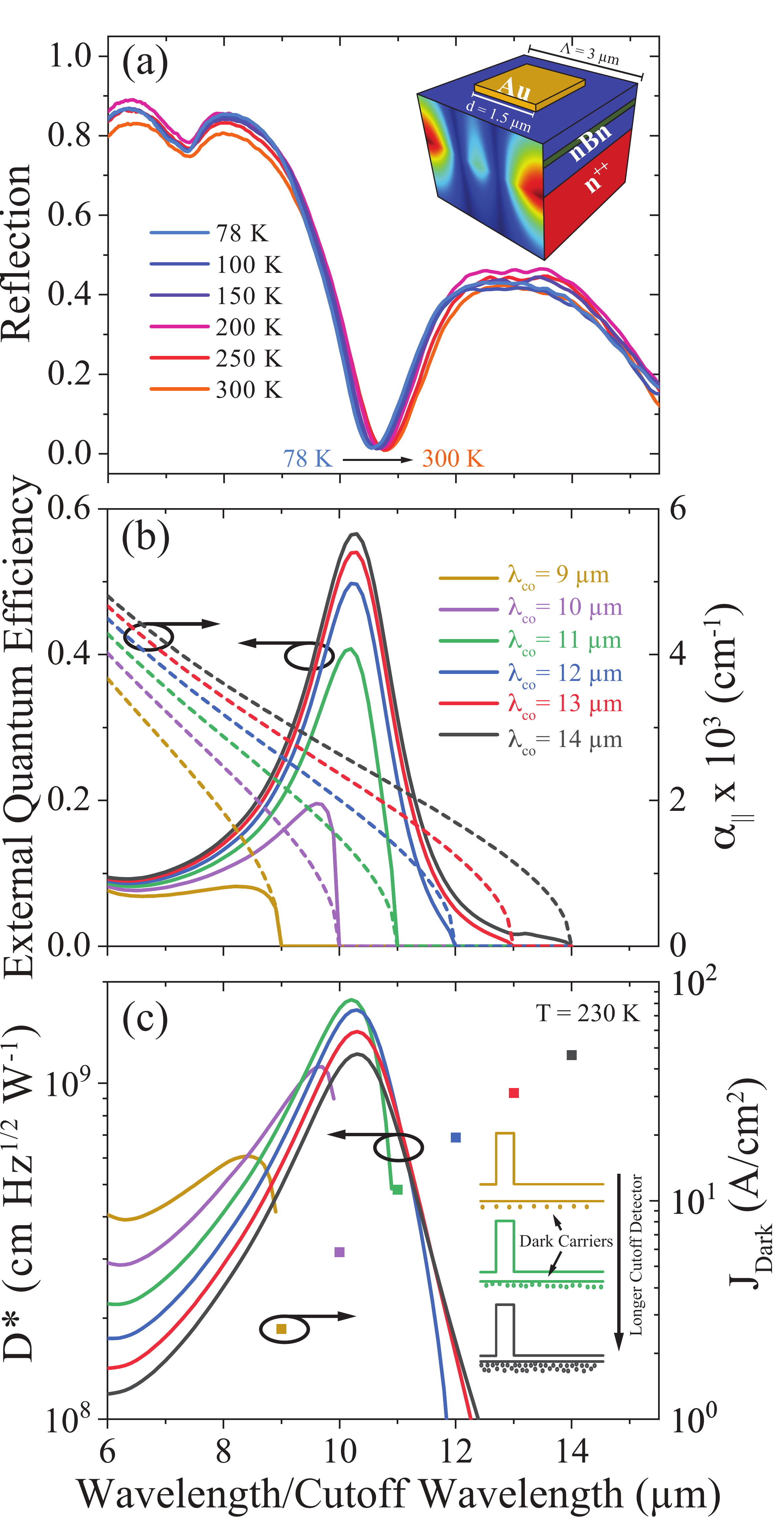}
  \caption{(a) Measured temperature-dependent reflection of the fabricated plasmonic ultra-thin detector from 78 K to 300 K. Note that the dip around \SI{10.4}{\um}, associated with coupling to a surface plasmon-polariton, shows a negligible spectral shift across the full temperature range. Inset is a schematic of one period of the ultra-thin plasmonic nBn detector and field profiles at resonance ($\lambda=\SI{10.4}{\um}$). (b) {Simulated transverse absorption coefficient (dashed lines) and quantum efficiency (solid lines) of the plasmonic detector as a function of absorber  cutoff wavelength (0\% cutoff).} As the absorber bandgap extends to longer wavelengths, diminishing returns in the peak external quantum efficiency are obtained. (c)  Calculated specific detectivity vs. wavelength (solid lines) and dark current vs. cut-off wavelength calculated using the ``Rule 07'' heuristic (scatter plot). As the detector cutoff is extended to longer wavelengths, specific detectivity initially increases, but then quickly declines due to a combination of added dark current and effectively constant EQE. The inset is a schematic depiction of band structure and minority carrier concentration for nBn T2SL detectors with increasing cutoff wavelength.}
  \label{fig:1}
\end{figure}

A particularly successful implementation of optically-enhanced T2SL detectors is the recently-demonstrated ultra-thin ($t\sim\lambda_0/30$) plasmonic detector\cite{Nordin2021Ultra-thinDetectors}, which realizes responsivities commensurate with much thicker traditional detectors ($t\sim\lambda_0$), while also demonstrating dark currents substantially below ``Rule 07''\cite{Tennant2008MBEHeuristic}, the heuristic for HgCdTe detectors and thus a useful benchmark for competing detector technologies. In such an architecture, an ultra-thin ($\sim311~nm$) T2SL detector structure is grown upon a heavily ($n^{++}$) doped semiconductor `ground-plane'. The interface between this (plasmonic) $n^{++}$ semiconductor material \cite{Law2012Mid-infraredMetals, Law2013EpitaxialPlasmonics} and the (dielectric) detector supports surface plasmon-polariton (SPP) modes at LWIR wavelengths.  Incident light is coupled into the tightly-bound SPP modes via a patterned 2D grating on the detector surface (Fig. \ref{fig:1}(a) inset). The tight confinement of these SPP modes allows for the absorber thickness, and thus dark current, to be drastically reduced.  

For the ultra-thin plasmonic detectors, the spectral position of the resonant plasmonic mode (for a given 2D grating geometry and detector thickness) is largely temperature-invariant [Fig. \ref{fig:1}(a)], while the effective bandgap of the T2SL decreases with increasing temperature. {Figure \ref{fig:1}(b) shows rigorous coupled wave analysis\cite{FromAnalysis} (RCWA) simulations of plasmonic detector external quantum efficiency (EQE) for absorber cutoff wavelengths from \SI{9}{\um} to \SI{14}{\um}. For these simulations, we use a complex uniaxial permittivity tensor for the T2SL \cite{Gautam2012IdentificationSpectroscopy} with $\epsilon_{||}$ obtained from our previous work\cite{Nordin2021Ultra-thinDetectors} and $\epsilon_{\perp} = 12.56$. As can be seen in Fig. \ref{fig:1}(b), the simulated absorption (EQE) rapidly increases as the cutoff of the absorber sweeps into the spectral position of the plasmonic mode, after which the absorption (EQE) saturates as the cutoff of extends past the plasmonic mode.}  Thus, for a detector with excellent low temperature performance, increasing the detector operating temperature offers little additional benefit to responsivity while incurring dramatic increases in dark current [Fig. \ref{fig:1}(c)], resulting in significant degradation of detector specific detectivity ($D^*$). However, if the effective bandgap of the absorber is tailored to extend, at elevated temperatures (T > 200 K), to the spectral position  of the plasmonic mode (as opposed to well-past this position on the low-energy side), one would expect to be able to realize the superior performance of the ultra-thin plasmonic T2SL detector, but at significantly elevated temperatures. 

Here we demonstrate the flexibility of the plasmonic T2SL detector architecture and, most importantly, the potential for plasmonic T2SL detectors to out-perform commercial near-room temperature thermoelectrically cooled HgCdTe detectors. {We achieve a substantial operating temperature and performance increase in the plasmonic detectors, relative to the initial demonstration of these devices\cite{Nordin2021Ultra-thinDetectors}, by co-designing the optical and electronic structure of the detector architecture. Optically, we leverage the relative temperature invariance of the plasmonic mode. Electronically, we replace the barrier superlattice with a wider bandgap ternary layer of AlAs$_{0.1}$Sb$_{0.9}$ to reduce thermionic emission dark current, adjust the absorber superlattice geometry for appropriate cutoff wavelength at high temperatures, and reduce the barrier layer thickness while increasing the absorber layer thickness to maximize EQE and minimize detector turn-on voltage. A detailed description of these adjustments and the detector design evolution is provided in the Supplemental Material.} We characterize our detector's electrical and optical response and compare the results to the ``Rule 07'' heuristic and commercial MCT detectors. Our detector, operated at 230 K, achieves external quantum efficiency as high as 22.8\% in only a 372 nm thickness absorber with dark currents $4\times$ lower than ``Rule 07'', {a factor of two better than our initial implementation\cite{Nordin2021Ultra-thinDetectors},} resulting in a peak specific detectivity of \SI{2.29E9}{\cm\Hz^{1/2}\watt^{-1}} at \SI{9.6}{\um}. Our plasmonic detector's specific detectivity is nearly an order of magnitude higher than commercial HgCdTe detectors operating at the same high temperature (230 K). The high operating temperature (HOT) plasmonic detector presented in this work thus offers a viable alternative to environmentally-problematic thermoelectrically cooled HgCdTe. 

The detectors are grown on an n-type doped GaSb substrate by molecular beam epitaxy in a Varian Gen-II system with effusion sources for gallium, indium, aluminum, and silicon, and valved cracker sources for arsenic and antimony.  Growth begins with a $\SI{200}{\nm}$ {Te doped (n-type \SI{1e18}{cm^-3})} GaSb buffer, above which is grown the plasmonic virtual substrate, comprising 123 periods of $n^{++}$ Si doped mid-wave infrared (MWIR) InAs/InAs\textsubscript{0.49}Sb\textsubscript{0.51} T2SL, with each period having 16.5 ML of InAs and 3.5 ML of InAs\textsubscript{0.49}Sb\textsubscript{0.51}. Following this plasmonic virtual substrate, {the detector structure is grown, beginning with 47 periods of the unintentionally doped (UID) absorber superlattice, with each period having 21.4 ML of InAs and 4.6 ML of InAs\textsubscript{0.49}Sb\textsubscript{0.51}.} {Above the absorber superlattice, 80 nm of ternary AlAs\textsubscript{0.1}Sb\textsubscript{0.9} (UID) is grown, which acts as a barrier layer}. {We then grow 7 periods of the contact layer T2SL (UID), which has the same composition as the absorber T2SL.} The $n^{++}$ plasmonic layer is grown with a doping concentration of $\sim \SI{5e19}{\cm^{-3}}$, and corresponding Drude parameters\cite{Law2012Mid-infraredMetals} of a \SI{5.5}{\um} plasma wavelength ($\lambda_p=2\pi c/\omega_{p} =\SI{5.5}{\um}$) and a \SI{1e13}{} rad/s scattering rate ($\gamma=\SI{1e13}{} $ rad/s). Note that while we use a T2SL as our plasmonic layer to simplify lattice-matching to the layers above and below the plasmonic film, the higher energy bandgap of the n$^{++}$ T2SL, combined with state-filling resulting from the high doping, allows this layer to simply be treated as a Drude metal\cite{Nordin2020All-EpitaxialResponsivity,Nordin2021Ultra-thinDetectors,Abhilasha_GMR}. %This is due to the combination of the MWIR cut-off and state filling from the high doping concentration (Burstein Moss shift)\cite{Moss1954TheAntimonide, Burstein1954AnomalousInSb}, which ensures that the loss in the $n^{++}$ layer comes from the Drude scattering term ($\gamma$) and not interband absorption.

\begin{figure}[!t]
  \includegraphics[scale=0.27]{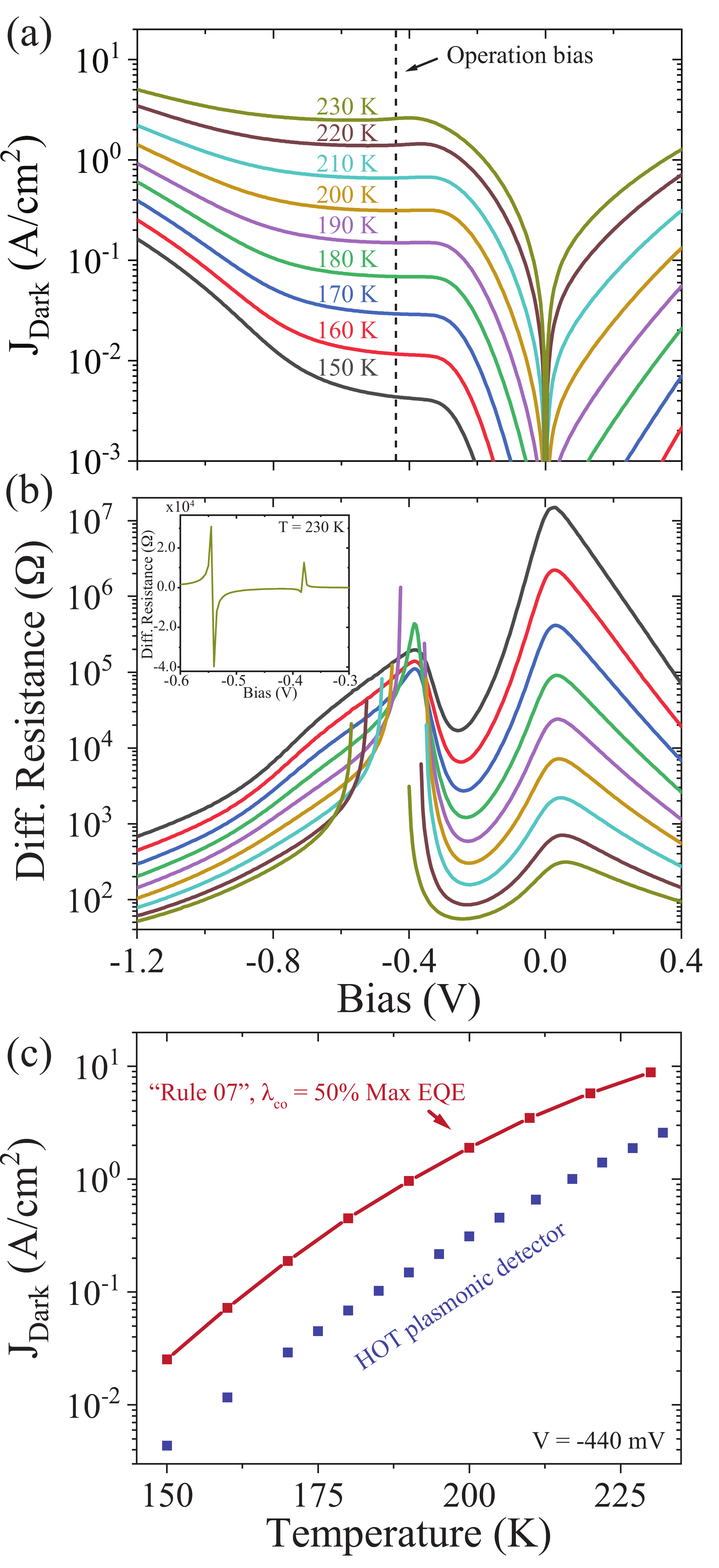}
  \caption{(a)  Temperature dependent dark current of the HOT plasmonic detector. For moderate biases (-370 to -600 mV), the dark current is relatively flat, characteristic of a diffusion-limited dark current. {(b) Temperature dependent differential resistance for the same temperature range show in (a). Negative differential resistance is observed for temperatures greater than 180 K. Inset shows the linear scale differential resistance for  230 K. (c)} {Dark current at a bias of -440 mV (blue)  as a function of temperature compared to a ``Rule 07'' with cutoff wavelength chosen by taking 50\% max EQE at each temperature (red).} At 230 K, where the HOT plasmonic detector has a cutoff wavelength of \SI{10.6}{\um}, the HOT plasmonic detector has a dark current $\sim 4 \times$ less than ``Rule 07''.}
  \label{fig:2}
\end{figure}

{Following epitaxial growth, the as-grown material is patterned into mesas of dimensions \SI{340}{\um} x \SI{540}{\um} using UV photolithography and a wet etch.  The mesas are etched to a depth of 750 nm in order to provide electrical isolation between mesa top contacts.  The detector top and bottom contacts, as well as the 2D metal gratings are then defined using UV photolithography and a metallization and lift-off process.  The metal layer stack is 3 nm of Pd, 50 nm of Ti, 50 nm of Pt, and 100 nm of Au, which is a standard shallow contact for narrow bandgap T2SL materials. Detectors are left unpassivated, however following lift-off the detectors are thoroughly cleaned in a bath of 90 °C AZ KWIKSTRIP followed by sonication in organic solvents (acetone, methanol, and isopropyl alcohol).}  For the HOT plasmonic detectors presented here, we use a 2D grating array with a period of \SI{3}{\um} and a 50\% duty cycle designed to couple incident LWIR light to SPP modes bound to the $n^{++}$/T2SL absorber interface. Following fabrication, detectors were mounted onto leadless chip carriers in a custom-designed temperature-controlled cryostat. Dark current measurements were taken with the detector blocked by a copper cold shield thermally connected to the cryostat cold-finger. Temperature dependent detector dark current as a function of applied bias (J-V) was measured using a Keithley 2460 low-noise source-meter and is shown in Figure \ref{fig:2}(a). 

From -300 mV to -600 mV the JV curves are effectively flat, suggesting a diffusion-limited dark current mechanism. Beyond -600 mV, the dark current is relatively temperature insensitive, suggesting the majority of the additional dark current at high biases is due to band-to-band tunnelling and/or mid-gap carrier generation. {Shown in Fig. \ref{fig:2}(b) is the temperature dependent differential resistance. For elevated temperatures (T$>185$K), we observe clear negative differential resistance (NDR) at the onset of diffusion limited behavior.} NDR in photodetectors has been analytically described\cite{White1986AugerStructures,White1987NegativeStructures} and observed in HgCdTe detectors\cite{Ashley1985NonequilibriumDetection} and MWIR III-V detectors \cite{Ashley1991OperationTechniques}, and has traditionally been attributed to depletion and therefore substantial Auger suppression in the detector's absorber. Given the ultra-thin absorber utilized in the plasmonic detectors,  the experimentally-observed NDR in our device could be attributed to a similar depletion mechanism and a commensurate Auger suppression. {Shown in Fig. \ref{fig:2}(c) is temperature dependent dark current at a bias -440 mV compared to a ``Rule 07'' where cutoff wavelength at each temperature is determined by the spectral position of 50\% of max EQE at that temperature} For all temperatures the plasmonic detector shows dark current well below ``Rule 07''. Specifically, at 230 K the HOT plasmonic detector possess a dark current $\sim 4 \times$ lower than ``Rule 07''. 

\begin{figure}[!t]
  \includegraphics[width=\columnwidth]{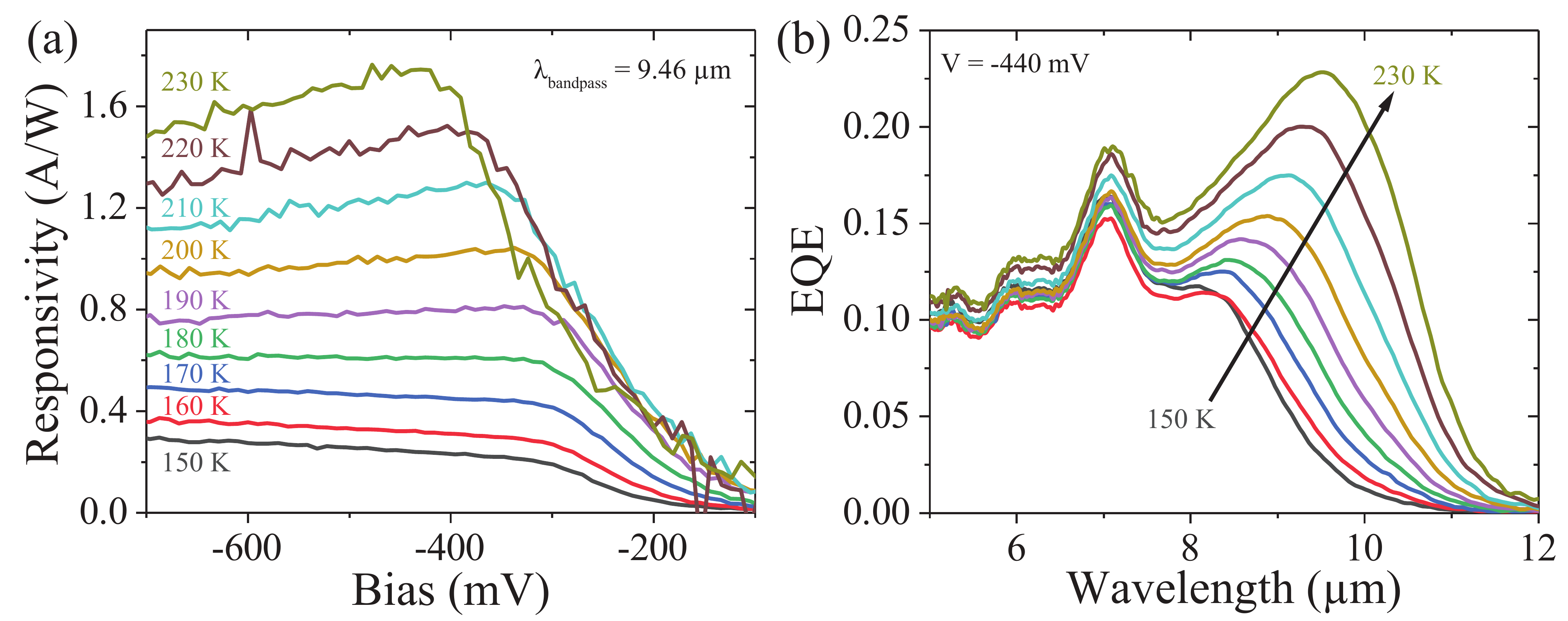}
  \caption{(a) Bias-dependent responsivity (at $\lambda_{ bandpass}=\SI{9.46}{\um}$) of the HOT plasmonic detector as a function of temperature. Responsivity increases substantially as a function of temperature due to the detector's cutoff wavelength shifting to longer wavelengths and therefore spectrally aligning with the plasmonic enhancement feature. (b) External quantum efficiency (EQE) spectra as a function of temperature of the HOT plasmonic detector. EQE, similar to responsivity, increases substantially at elevated temperatures. }
  \label{fig:3}
\end{figure}

The fabricated detector responsivity was measured using a calibrated blackbody source; details of the measurement and spectral response measurements are described in previous work\cite{Nordin2021Ultra-thinDetectors}. Shown in Fig. \ref{fig:3}(a) is the bias-dependent responsivity as a function of temperature. As anticipated for barrier detectors, the responsivity peaks at the onset of  diffusion-limited behavior we observe in Fig. \ref{fig:2}(a) (relatively flat J-V). {However, this saturation occurs at relatively high biases, near -300 mV at low temperatures and as high as -440 mV at 230 K, suggesting the valence band offset between the barrier and the T2SL is nonzero. Future iterations of these devices could lower this turn-on bias by implementing an optimized barrier layer.} Additionally, the responsivity increases significantly at elevated temperatures, which is expected behavior resulting from the detector absorber's cutoff wavelength redshifting into the plasmonic mode and thus enabling the enhanced absorption of SPPs by the detector's absorber. However, to confirm this anticipated behavior, we need the detector's EQE spectrum to accurately compare to the simulations show in Fig. \ref{fig:1}(b). Shown in Fig. \ref{fig:3}(b) is the measured EQE as a function of temperature at -440 mV. As expected, the EQE increases substantially as the detector cutoff redshifts into the plasmonic mode hosted at the $n^{++}$/T2SL absorber interface. {The substantial increase, and then relative saturation, in EQE that we observe in our devices is nearly identical to the simulated EQE in Fig. \ref{fig:1}(b), especially for the $\lambda_{co} =$ \SI{9}{\um} (yellow) to \SI{11}{\um} (green) lines. However, we do not reach the nearly 50\% EQE in Fig. \ref{fig:1}(b) for the simulated $\lambda_{co} =$ \SI{11}{um} or \SI{12}{\um} detectors, suggesting that the cutoff wavelength of the present detectors could be made slightly longer at 230 K and the increase in EQE may balance the additional dark current. } {An additional feature at \SI{7}{\um} is present, and we attribute this feature to a TE-polarized guided-mode-resonance. Our 2D simulations in Fig. \ref{fig:1}(b) do not include this feature because they only consider the (in-plane) TM-polarized absorption.} 

\begin{figure}[!t]
  \includegraphics[scale = 0.31]{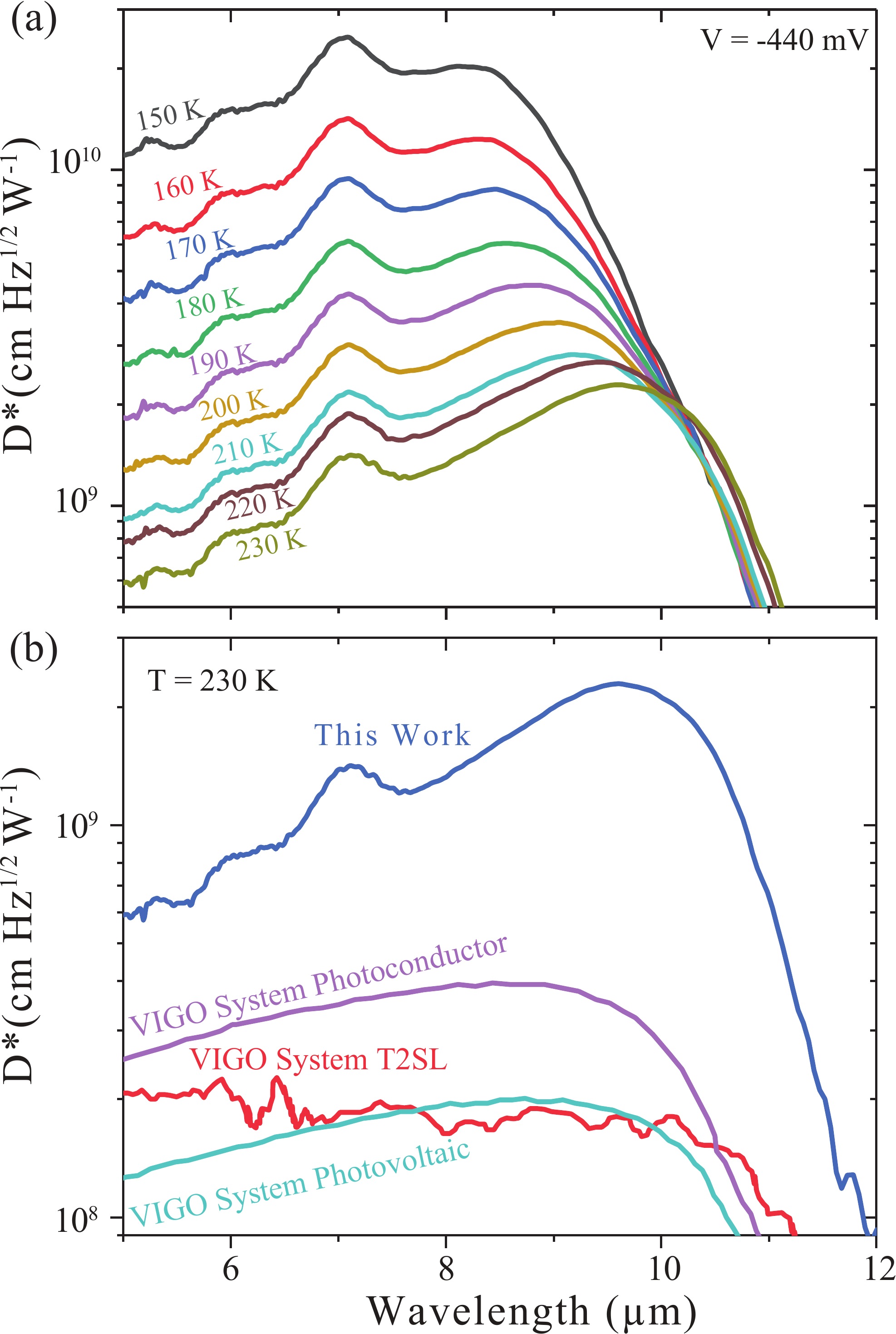}
  \caption{(a) The HOT Plasmonic detector's spectral resolved specific detectivity as a function of temperature. Specific detectivity stays relatively constant as a function of temperature due to the increase dark current being balanced by the increasing responsivity. (b) Specific detectivity of commercial detectors from VIGO system compared to the HOT plasmoinc detector. }
  \label{fig:4}
\end{figure}

We estimate specific detectivity with the expression \begin{math}
    D^{*} = R_i  \sqrt{\frac{A}{2 q \abs{I} + 4 k_b T/R}}
\end{math}, using the measured detector dark current $(I)$ and responsivity ($R_i$), with $A$ the mesa area, $q$ the electronic charge, Boltzmann's constant $k_b$ , $T$ the detector temperature, and $R$ the differential resistance. The temperature-dependent estimated spectral specific detectivity at a bias of -440 mV is shown in Fig. \ref{fig:4}(a). Between 150 K and 230 K the peak specific detectivity only decreases by an order of magnitude. {Above 230 K the low-noise preamplifier (SRS 570) used for responsivity and spectral response measurements overloads, so we cannot provide estimated specific detectivity for these higher temperatures. Note that unlike the simulations presented in Fig. \ref{fig:1}(c), the experimentally-obtained estimated peak $D^*$ decreases monotonically with temperature. The simulations in Fig. \ref{fig:1}(c) show $D^*$ at a fixed temperature for changing cut-off wavelengths (different superlattices). Our experimental results are sweeping both cutoff wavelength and temperature, the latter of which accounts for the additional, and unavoidable, decrease in $D^*$.} We attribute the relatively small reduction in specific detectivity to the exponentially increasing dark current being partially compensated by the increasing EQE as the absorber band edge moves through the optical resonance of the device. To put the current work in context, we compare the results from our HOT plasmonic detector to commercial detectors by plotting our estimated specific detectivity against VIGO system's thermoelectrically cooled (230 K) MCT detectors\cite{VIGOPC,VIGOPV}, both photoconductive and photovoltaic, and VIGO System's new line of thermoelectrically cooled (230 K) photoconductive T2SL detectors\cite{VIGOT2SL}. The HOT plasmonic detector presented in this work shows peak specific detectivities of \SI{2.29E9}{\cm\Hz^{1/2}\watt^{-1}} at 230 K, while the commercial detectors only achieve specific detectivities of \SI{1E8}{\cm\Hz^{1/2}\watt^{-1}} to \SI{4E8}{\cm\Hz^{1/2}\watt^{-1}}. 

The plasmonic LWIR detectors show high performance when the absorber cut-off wavelength sits just beyond the wavelength of the plasmonic resonance, but performance degrades quickly with increasing temperature due to increasing dark current resulting from the decrease in the absorber bandgap.  The substantial design flexibility offered by the use of designer epitaxial metals and quantum-engineered device active regions allows for the careful tailoring of the absorber for strongly-enhanced LWIR detection at a given temperature.  Specifically, by engineering the electronic and optical structure of our plasmonic detectors, we realize a LWIR detector, operating at 230 K, which out-performs commercial detectors at the same temperature.  Moreover, we do so in a III-V material system, which benefits from a mature and well-established semiconductor fabrication and processing infrastructure, and which does not suffer from the environmental concerns and restrictions associated with the HgCdTe alloys. Although these results are a major leap forward in LWIR detector development, there is still a real need to push the operation of high performance infrared detectors to room temperature. A further increase in the operation temperature of the ultra-thin plasmonic detectors will likely require leveraging  Auger suppression techniques, perhaps via carefully engineered superlattices \cite{Meyer1995Type-IIIN,MuhowskiAuger2020} and/or by implementing novel electronic device architectures\cite{White1986AugerStructures,law19}. The demonstrated detectors highlight only one of a range of optical designs incorporating integrated plasmonic layers. There exists considerable potential for LWIR detector architectures with enhanced functionality, such as ultra-fast detection, or spectrally tunable/switchable detection which could be enabled by different plasmonic optoelectronic architectures \cite{LRSPP2021Nordin}. The detectors demonstrated here are thus an important step towards room temperature devices, as well as device architectures with the potential to extend the limits of LWIR detection and detector functionality.

In summary, we demonstrate a HOT long-wavelength infrared detector plasmonic detector which out-performs commercial HgCdTe detectors. We achieve this superior performance by carefully tailoring the optical and electronic band structure of our detector for enhanced LWIR absorption at elevated temperatures. The plasmonic detectors at 230 K possess EQEs of 22.8\% with dark currents $4\times$ lower than ``Rule 07'', and thus peak specific detectivity of \SI{2.29E9}{\cm\Hz^{1/2}\watt^{-1}}. The peak specific detectivity achieved by the plasmonic detectors is nearly an order of magnitude larger that that of commercial MCT detectors operated at the same temperature. The work presented here demonstrates that not only are the plasmonic T2SL detectors a viable replacement for single-element HgCdTe detectors, but in fact out-perform these state-of-the-art commercial detectors on the key figures of merit for detector performance. 

\section*{Supplementary Material}
See the supplementary material for x-ray diffraction measurements of as-grown device material. 

\begin{acknowledgments}
DW and LN gratefully acknowledge support from the National Science Foundation (ECCS-1926187). This work was performed in part at the University of Texas Microelectronics Research Center, a member of the National Nanotechnology Coordinated Infrastructure (NNCI), which is supported by the National Science Foundation (grant ECCS-2025227).
\end{acknowledgments}

\section*{Data Availability}
The data that support the findings of this study are available from the corresponding author upon reasonable request.

\bibliography{references}% Produces the bibliography via BibTeX.

\end{document}

\title{High operating temperature plasmonic infrared detectors \\ Supplementary Material}
\maketitle

\section*{Detector design}
% Force line breaks with \\
The first published demonstration of the plasmonic detectors\cite{Nordin2021Ultra-thinDetectors}  operated up to 195 K, with EQE curves indicating efficient absorption of incident light coupled into the plasmonic mode, a result of the detector cutoff wavelength being much longer than the wavelength of the resonant coupling to the surface plasmon polariton. As thoroughly discussed in the main text, extending the cutoff wavelength beyond the wavelength of plasmonic mode provides little additional EQE, while contributing substantial dark current to the device operation. In this section we describe the design evolution of these detectors, from initial experimental demonstrations, to the devices of Ref. \cite{Nordin2021Ultra-thinDetectors}, to the detectors of the current work. Shown in Table \ref{table:1} is a list of representative detectors and their associated growth parameters and operating properties.  Because the plasmonic cavity modulates the device photoluminescence, the cut-off wavelength of our detectors is instead extracted from the detector response. For all of the detectors in Table \ref{table:1}, we provide two cut-off wavelengths, the first corresponding to the spectral position where we see $50\%$ of maximum EQE (EQE\textsubscript{max}), the second where EQE drops to $1\%$.  Though the former is most frequently used in traditional detector characterization, it is less useful for our plasmonic detectors, because for effective bandgaps smaller than the plasmonic resonance, $50\%$ of EQE\textsubscript{max} will always be tied to the position of the plasmonic resonance, not the material bandgap.  The latter ($<1\%$ EQE\textsubscript{max}) provides a more accurate estimate of the true effective band-gap of the superlattice absorber material.  

\begin{table}[h!]
\centering
\begin{tabular}{||c c c c c c c c c||}
 \hline
 $T_{max} (K)$ & 
 $t_{InAs}$ (ML) & 
 $t_{InAsSb}$ (ML)&  $\lambda_{50\% \text{ Max EQE}}$ ($\mu$m) &
 $\lambda_{1\% EQE}$ ($\mu$m) &Barrier Type & 
 $t_{Barrier}$ (nm) &
 $V_{op}$ (mV) & 
 Peak EQE (\%)\\ [0.5ex] 
 \hline\hline
 170 & 
 33 & 
 7 & 
 11.4 &
  16.4 & 
 SL\cite{Haddadi2017} & 
 148 & 
 -200 & 
 47.5\\ 
 \hline
 
 185 &
 28.9 &
 6.1 & 
 11.5 & 
 14.6 & 
 SL\cite{Nordin2021Ultra-thinDetectors} & 
 146 & 
 -175 & 
 43.3\\
 \hline
 
 195 & 
 24.7 & 
 5.3 & 
 11.4 &
 13.5 &
 SL\cite{Nordin2021Ultra-thinDetectors} &
 146 &
 -160 &
 39.0\\
 \hline
 
 205 & 
 24.7 & 
 5.3 & 
 11.3 & 
  13 & 
 AlAs$_0.1$Sb$_0.9$ & 
 146 &
 -1025 &
 37.6\\
 \hline
 
 230 (this work) & 
 21.4 & 
 4.6 & 
 10.6 & 
 11.6 & 
 AlAs$_0.1$Sb$_0.9$ & 
 80 & 
 -440 & 
 22.8\\
 \hline
\end{tabular}
\caption{Table of representative plasmonic detectors with associated growth parameters and operating properties, including operating temperature, superlattice period geometry, cutoff wavelength, barrier type, and barrier thickness.  }
\label{table:1}
\end{table}

The first three generations of plasmonic detectors ($T_{max}=170,~185,~195$K) used superlattice absorbers with decreasing period ($40,~35,~30$ ML) and superlattice (SL) barriers. Decreasing the absorber SL period blue-shifts the effective bandgap (as seen in the $\lambda_{1\% EQE_{max}}$ column), with little to no effect on the position of the absorption peak, and subsequent position of $\lambda_{50\%~EQE_{max}}$. Increasing the SL effective bandgap results in a manageable decrease in peak EQE (from $47.5\%$ to $39.5\%$), but significantly reduces dark current, resulting in an increase in operating temperature, with the device having the shortest period SL absorber reaching a temperature accessible by thermoelectric coolers\cite{Nordin2021Ultra-thinDetectors}.  The fixed $\lambda_{50\%~EQE_{max}}$ for these three detectors suggests that the effective bandgaps ($\lambda_{1\%~EQE}$) are still long enough to absorb significant light at the wavelength of the plasmonic mode, suggesting that the cutoff wavelength ($\lambda_{1\%~EQE}$) could be further shortened. Additionally, concerns about potential thermionic emission over the superlattice barrier at these high temperatures ($J_{TE} \sim T^2$) motivated an adjustment to the barrier design. 

\begin{figure}
 \begin{center}
  \includegraphics[width = \columnwidth]{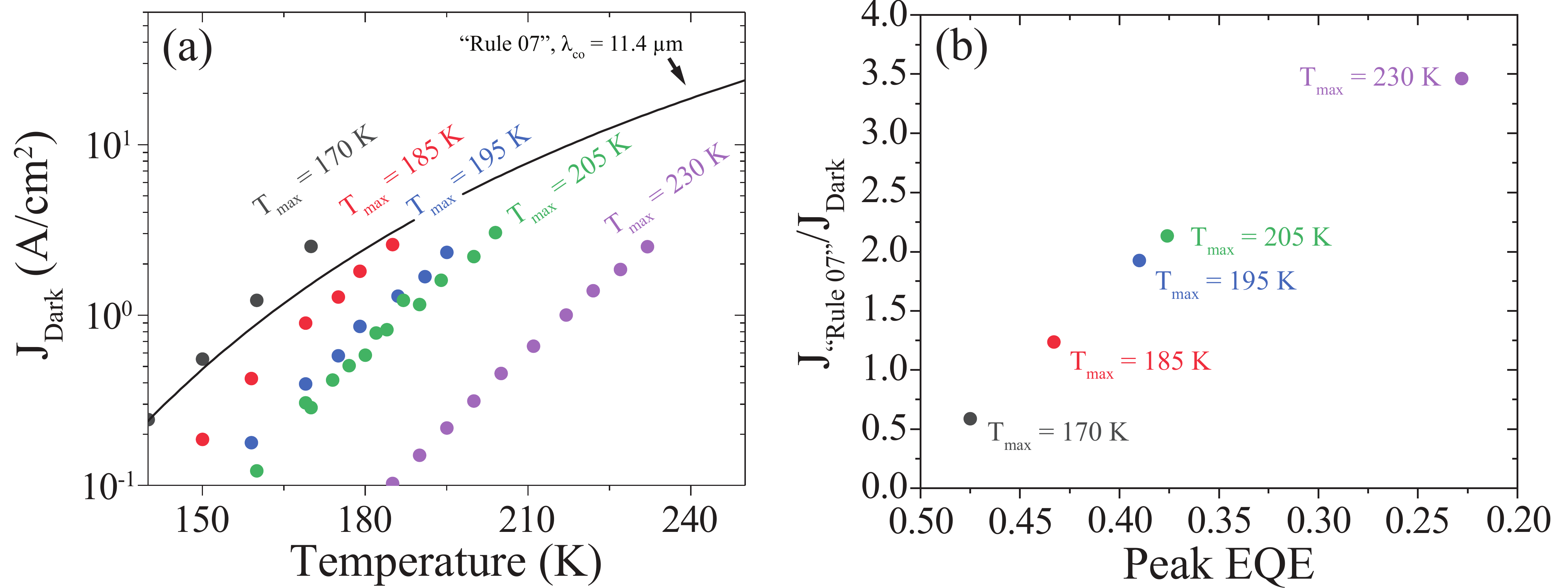}
  \caption{(a) Dark current as a function of temperature for the plasmonic detectors of Table \ref{table:1}. ``Rule 07'' is plotted for a cutoff wavelength of \SI{11.4}{\um}. Excluding the 170 K detector, all generations operate with dark current lower than  ``Rule 07''. (b) Dark current improvement relative to  ``Rule 07'' versus peak EQE. As the detector cutoff wavelength is shortened the dark current improves relative to Rule 07, at the price of decreasing EQE.}
  \label{SM1}
  \end{center}
\end{figure}

A detector was grown identical to the $T_{max}=195$ K detector, but with a wider bandgap ternary AlAs$_{0.1}$Sb$_{0.9}$ barrier, chosen following the Jet Propulsion Lab's demonstration of mid-wave infrared detectors with ternary AlAsSb barriers\cite{Ting2018Mid-wavelengthArray} that show low turn-on biases (-100 to -200 mV) at elevated temperatures (150 K). Changing the barrier to the ternary AlAsSb results in an improvement in operating temperature to 205 K. However, these detectors possessed an unreasonably high turn on bias (nearly -3 V at 77 K and -1 V at 205 K), far greater than any valence band offset observed in the literature for the Ga-free T2SL system\cite{Ting2018Mid-wavelengthArray}. 

The uncharacteristically high turn-on biases in the 205 K detector suggest that the new ternary barrier is n-type unintentionally doped (UID) at a relatively high ($\sim$\SI{1E16}{\cm^{-3}}) concentration, which would induce a diffusion barrier to carrier collection. Reduction in the voltage needed to overcome this diffusion barrier could be achieved with either a thinner barrier layer or by counter-doping the ternary layer to reduce UID concentration. However, were the sample to be counter-doped at a level even slightly greater than the UID of the barrier, the barrier would become p-type, and field would be dropped in the absorber, likely resulting in a non-negligible depletion region generation-recombination dark current. To avoid this, we instead grew a thinner AlAs$_{0.1}$Sb$_{0.9}$ barrier layer (80 nm) and a thicker (372 nm) shorter cutoff wavelength absorber (the detector of the main text). 

Figure \ref{SM1}(a) shows dark current density as a function of temperature for the detectors of Table \ref{table:1}. The $T_{max} = 230 $ K detectors possess the lowest dark current of all detectors, nearly an order of magnitude lower than the $T_{max} = 195$ K detectors, likely a consequence of the wider bandgap absorber and the increased bandgap ternary barrier. However, detector performance is both a function of dark current and EQE, and the increase in bandgap is clearly affecting absorption efficiency at the plasmonic resonance, reducing the peak EQE of the $T_{max}=230$ K detector. Figure \ref{SM1}(b) shows the dark current relative to ``Rule 07'' versus peak EQE, for the detectors in Table \ref{table:1}. The clear trade-off in detector design can be observed: as the cutoff wavelength decreases, our operation temperature is increased and our dark current relative to ``Rule 07'' is improved, while the measured peak EQE monotonically decreases. 

\section*{Structural Characterization}
     \begin{figure}[!h]
  \includegraphics[scale = 0.5]{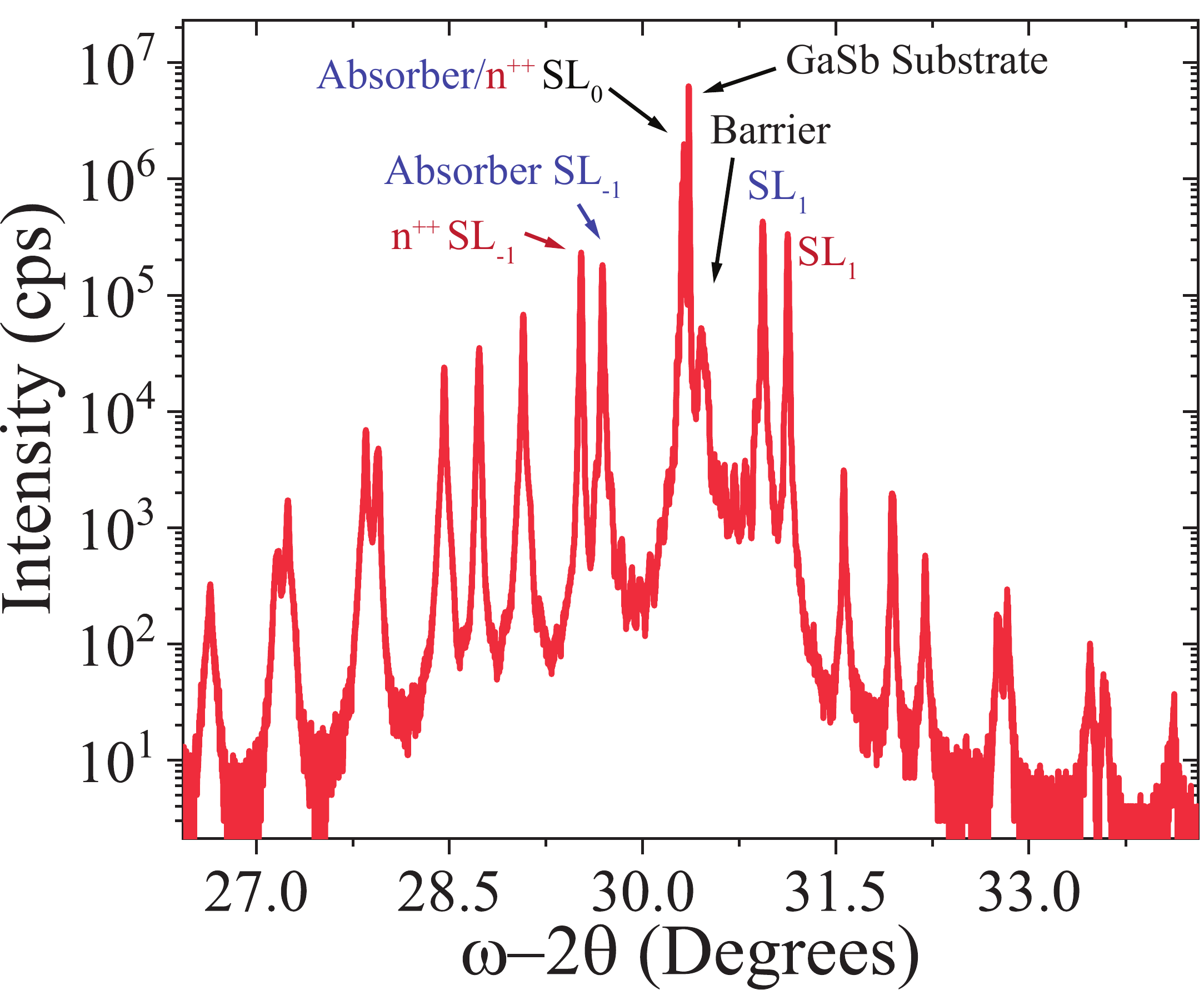}
  \caption{ X-ray diffraction (004) $\omega - 2\theta$ scan of the high-operating-temperature devices. The substrate, AlAsSb barrier layer, 26 ML absorber superlattice, and 20 ML $n^{++}$ superlattice are noted on the plot. The substrate peak and absorber/$n^{++}$ zeroth order superlattice peak are nearly overlapped, suggesting excellent lattice matching.}
  \label{SM2}
\end{figure}

\bibliography{references}% Produces the bibliography via BibTeX.